\documentclass[11pt]{article}

\usepackage{amsmath,amssymb}
\usepackage{slashed}
\usepackage{xcolor} 
\usepackage{graphicx}
\usepackage{dcolumn} 
\usepackage{bm} 
\usepackage{epsfig}
\usepackage{epstopdf}
\usepackage{grffile}
\usepackage{color}
\usepackage{colordvi}
\usepackage{amsmath,amssymb}
\usepackage{rotating}
\usepackage{lscape}
\usepackage{cite}
\usepackage{float}
\usepackage{hyperref}

\def\Re{{\cal R \mskip-4mu \lower.1ex \hbox{\it e}\,}}
\def\Im{{\cal I \mskip-5mu \lower.1ex \hbox{\it m}\,}}

\def\tev{\,{\ifmmode\mathrm {TeV}\else TeV\fi}}
\def\gev{\,{\ifmmode\mathrm {GeV}\else GeV\fi}}
\def\mev{\,{\ifmmode\mathrm {MeV}\else MeV\fi}}

\begin{document}

\begin{center}

\vspace*{15mm}
\vspace{1cm}
{\Large \bf Measuring anomalous WW$\gamma$ and  t$\bar{\text{t}}\gamma$  couplings using top+$\gamma$ production at the LHC}

\vspace{1cm}

{\bf  Seyed Mohsen Etesami, Sara Khatibi,  and Mojtaba Mohammadi Najafabadi }

 \vspace*{0.5cm}

{\small\sl School of Particles and Accelerators, Institute for Research in Fundamental Sciences (IPM) P.O. Box 19395-5531, Tehran, Iran } \\

\vspace*{.2cm}
\end{center}

\vspace*{10mm}

%
%
\begin{abstract}\label{abstract}
We consider the electroweak production of a top quark in association with a photon at the LHC
to probe the electroweak top quark couplings (t$\bar{\text{t}}\gamma$) as well as the triple gauge boson couplings (WW$\gamma$).
The study is based on the modifications of the t$\bar{\text{t}}\gamma$ and WW$\gamma$ interactions
via heavy degrees of freedom in the form of dimension-six operators which we add to the standard model Lagrangian.
A binned angular asymmetry in single top quark plus photon events and cross section ratio are proposed to probe the anomalous
 t$\bar{\text{t}}\gamma$ and WW$\gamma$
couplings.  It is shown that the proposed angular asymmetry can distinguish anomalous
t$\bar{\text{t}}\gamma$, WW$\gamma$ couplings from the standard model prediction
and yield a great sensitivity.
\end{abstract}

\vspace*{3mm}



\newpage


\section{Introduction}\label{Introduction}

The standard model (SM) of particle physics has been found to be prosperous in explaining
the strong and electroweak interactions. 
However,  there are unanswered questions concerning possible
SM extensions that incorporate new particles and new interactions.
Studying top quark interactions and the electroweak gauge bosons self-interactions 
could provide applicable information in probing the extensions of the SM. 
As a result,  precise measurements of the top quark interactions and the SM gauge boson self-couplings are necessary since
any deviation from the SM forms and values would be indicative of new physics beyond the SM.
Anomalous triple gauge boson couplings and the top quark interactions have been  extensively studied in the literature, 
see for example  \cite{a1, a2, a3, a4, a5, a6, a7, htt, mmn, a8, a9, a10,a11,a12,a13,a14,a15,a16,a17,a18,a19,a20,a21,a22,a23,a24,a25, a26,a27,a28,a29,a30}
 and references therein.

A relevant approach in describing possible new physics effects is a 
model independent approach based on an effective field theory at low energy.
In such an approach, all the heavy degrees of freedom are integrated out leading to obtain
the effective interactions among the SM particles.  
This is justified due to the fact that the related observables have not shown any significant deviation from the SM predictions  so far.
These effective couplings are suppressed by the inverse powers
of the new physics scale $\Lambda$. The 
effective Lagrangian is required to satisfy the SM local symmetry $SU(3)_{C} \times SU(2)_{L} \times U(1)_{Y}$.
With the requirement of lepton and baryon number conservation,  the Lagrangian takes the following form \cite{efff1,eff2,eff3}:
\begin{eqnarray}
\mathcal{L}_{eff} = \mathcal{L}_{SM} + \sum_{i}\frac{c_{i}O_{i}}{\Lambda^{2}} + h.c.,
\end{eqnarray}
where $O_{i}$ are the gauge invariant operators of dimension-six and $c_{i}$ are the corresponding dimensionless coefficients. 
A list of dimension-six operators has been provided in \cite{efff1,eff2,eff3, efflag1 }. 
Recently, discussions on the validity of the effective field theory  extension of the SM with 
dimension-six operators and the fact that the validity  range of the effective theory  cannot  be  determined  just based on  
the low energy information have been provided in \cite{adam}.

The contributions from 
dimension-six operators including the SM coupling to the t$\bar{\text{t}}\gamma$ vertex is parameterized as follows \cite{eff2}:
\begin{eqnarray}\label{eff1}
\mathcal{L}_{t\bar{t}\gamma} = -eQ_{t}\bar{t}\gamma^{\mu}tA_{\mu}-ie\bar{t}\frac{\sigma_{\mu\nu}q^{\nu}}{2m_{t}}\Big(\kappa+i\bar{\kappa}\gamma_{5} \Big)tA^{\mu},
\end{eqnarray}
where the top quark charge and mass are denoted by $Q_{t}e$ and $m_{t}$, respectively.
The CP even parameter $\kappa$ and 
CP odd parameter $\bar{\kappa}$ are related to the top quark anomalous magnetic ($a_{t}$)
 and electric ($d_{t}$) dipole  moments via the following relations:
\begin{eqnarray}
\kappa = Q_{t} a_{t}~,~ \bar{\kappa}= \frac{2m_{t}}{e}d_{t}.
\end{eqnarray}

There two operators which contribute to  the top quark anomalous magnetic 
 and electric dipole moments \cite{eff2}:
 \begin{eqnarray} \label{op}
 \mathcal{O}^{33}_{uB\phi} =\bar{q}_{L3}\sigma^{\mu\nu}t_{R}\tilde{\phi}B_{\mu\nu} + h.c. ~~{\rm and}~~ 
  \mathcal{O}^{33}_{uW} =\bar{q}_{L3}\sigma^{\mu\nu}\tau^{a}t_{R}\tilde{\phi}W^{a}_{\mu\nu} + h.c.
 \end{eqnarray}
Based on the parameterization of Eq.\ref{eff1} and using the operators introduced in Eq.\ref{op}
, one finds:
\begin{eqnarray}
\kappa &=& \frac{2\sqrt{2}}{e}\frac{vm_{t}}{\Lambda^{2}}{\rm Re}[s_{W}C^{33}_{uW}+c_{W}C^{33}_{uB\phi}], \nonumber \\
\bar{\kappa} &=& \frac{2\sqrt{2}}{e}\frac{vm_{t}}{\Lambda^{2}}{\rm Im}[s_{W}C^{33}_{uW}+c_{W}C^{33}_{uB\phi}].
\end{eqnarray}
where $v = 246$ GeV and $s_{W}$ is the sine of Weinberg angle.
The prediction of the SM for the top quark anomalous magnetic dipole moment is $a_{t} = 0.02$
 which is corresponding to $\kappa = 0.013$ \cite{aat}.  The CP violating electric dipole moment $d_{t}$ 
appear at three-loop level and is arising from the complex elements of the CKM matrix. It is found to be at the order 
of $d_{t} < 10^{-30}~e.$cm corresponding to $\bar{\kappa} < 5.7\times 10^{-14}$ \cite{ddt}.
There are indirect constraints on the top quark magnetic and electric dipole moments from the 
b-quark rare decays $b\rightarrow s\gamma$ and the semi-leptonic b-quark decays \cite{bsg1,bsg2}.
The electric dipole moment, $d_{t}$, can also be constrained using the upper limit 
on the neutron electric dipole moment which was found to be 
$d_{t} <  3 \times 10^{-15} ~e.$cm \cite{neutron}.
The electric and magnetic dipole moments have been also probed using the direct 
$pp\rightarrow t\bar{t}\gamma$ production at the Tevatron and LHC. The combination of direct probe and 
the related b-quark decays leads to limits $a_{t} \in [-3,0.45]$ and $d_{t} \in [-0.29,0.86]\times 10^{-16}e.$cm \cite{bsg2}.
Indirect constraint on the top quark electric dipole moment coming from the ThO electric dipole moment
 measurement has been found to be $5\times 10^{-20}e.$cm \cite{a27}.

In \cite{tgamma},  the sensitivity of the single top quark production in association with a photon
to the anomalous electric and magnetic dipole moments of the top quark has been examined. 
An analysis on the several kinematic distributions of this process leads to constraints
 $a_{t} \in [-0.38,0.39]$ and $d_{t} \in [-0.15,0.15]\times 10^{-16}e.$cm at the LHC using 300 fb$^{-1}$ of integrated luminosity.

The dimension-six gauge invariant operators also contribute to the WW$\gamma$ coupling.
Under the assumption of charge conjugation and parity invariance,  the most general effective Lagrangian 
has the following form \cite{ww1,ww2}:
\begin{eqnarray}\label{eff2}
\mathcal{L}_{WW\gamma} = -ie\Big(W_{\mu\nu}^{\dagger}W^{\mu}A^{\nu}-W_{\mu}^{\dagger}A_{\nu}W^{\mu\nu}\Big)+i\kappa_{\gamma}
W^{\dagger}_{\mu}W_{\nu}F^{\mu\nu}+\frac{i\lambda}{m_{W}^{2}}W^{\dagger}_{\alpha\beta}W^{\beta}_{\delta}F^{\delta\alpha},
\end{eqnarray}
where $m_{W}$ is the $W$ boson mass, $W_{\mu\nu} = \partial_{\mu}W_{\nu}-\partial_{\nu}W_{\mu}$. In the SM, at tree level $\kappa_{\gamma}=1$
and $\lambda=0$.  At low energies, models with new heavy particles can effectively generate non-zero 
values for the anomalous triple gauge boson 
couplings $\Delta\kappa_{\gamma},\lambda$. 
These anomalous couplings $\lambda$ and $\Delta\kappa_{\gamma}$ (defined as $\kappa_{\gamma}-1$) 
have been probed indirectly using rare b-quark decay ($b\rightarrow s\gamma$) \cite{a13} and directly at colliders \cite{a2,a4}.
At the LHC, W$\gamma$ production has been used to probe the anomalous WW$\gamma$ couplings.
The $95\%$ CL limits on the anomalous couplings have been found to be $\Delta\kappa_{\gamma} \in [-0.38,0.29]$ and 
$\lambda\in[-0.050,0.037]$ from the CMS collaboration using 5 fb$^{-1}$ of 
proton-proton collisions at the center-of-mass energy of 7 TeV \cite{cmswg}. The ATLAS collaboration limits
at the $95\%$ CL have been found to be $\Delta\kappa_{\gamma} \in [-0.41,0.46]$ and 
$\lambda\in[-0.065,0.061]$ with 4.6 fb$^{-1}$ of 7 TeV data \cite{atlaswg}.
In addition to the above results from the CMS and ATLAS collaborations, 
the anomalous triple gauge boson couplings WW$\gamma$ have been studied at LEP \cite{lep1} and Tevatron \cite{tev1}.
The anomalous triple gauge boson couplings have been also confined 
by combining LEP data and the Higgs signal-strength data measured at the LHC experiments \cite{falkowski,a29}.

The aim of this paper is to  explore the possibility of constraining the top quark dipole moments 
as well as the anomalous triple gauge boson couplings WW$\gamma$ at the LHC
through photon radiation in single top events in t-channel mode. 
We concentrate on the leptonic decay mode with $l=e,\mu$ and construct an angular asymmetry 
in single top quark plus photon events to study the anomalous t$\bar{\text{t}}\gamma$ and WW$\gamma$
couplings.  We also consider the normalized cross section $\sigma_{tj\gamma}/\sigma_{tj}$ as a
 function of the anomalous couplings to set limits on those parameters. 
 
The paper is organized as follows. In Section \ref{xsection}, single top quark production
 in association with a photon is introduced. In Section \ref{xr} the normalized cross section $\sigma_{tj\gamma}/\sigma_{tj}$
 is  suggested and examined to explore the anomalous couplings  t$\bar{\text{t}}\gamma$ and WW$\gamma$.
In Section \ref{asymmetry}, a binned angular asymmetry which increase the sensitivity to possible new physics effects is proposed. 
Finally, in Section \ref{summary}, the summary and conclusions are given.

\section{Single Top Quark Production in Association with a Photon}
\label{xsection}

At the LHC within the SM framework, single top quarks in association with a photon 
can be produced through three separate channels. These channels can be categorized based on the way of
involvement of the W boson in the process. These processes are called  t-, s- and tW-channels. 
In the t-channel process, the top quark is produced via the exchange of a virtual and space-like
W boson. The involved W boson in s-channel top+$\gamma$ production is virtual and  time-like while in the 
tW-channel the involved W boson is a real W boson. The t-channel process has the largest production rate 
at the LHC.

\begin{figure}[htb]
\begin{center}
\vspace{1cm}
\resizebox{0.6\textwidth}{!}{\includegraphics{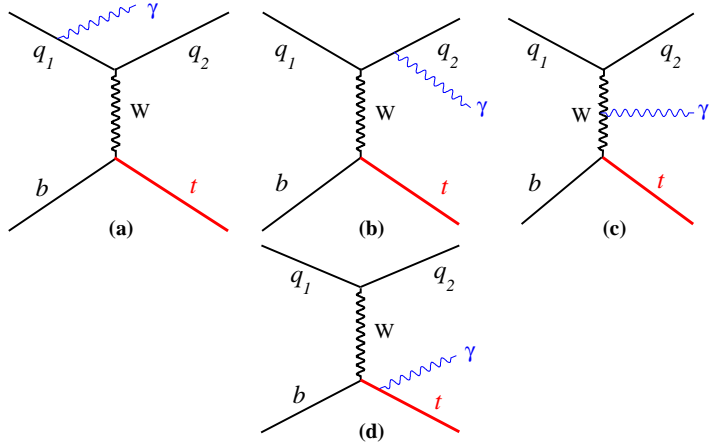}}  
\caption{  Representative leading order Feynman diagrams for the process of $tj\gamma$ production. }\label{feynman}
\end{center}
\end{figure}

We explore the potential of the LHC for probing the top quark electric and magnetic dipole moments as well as 
the triple gauge boson coupling WW$\gamma$ through photon radiation in single top events in the t-channel mode.
The calculations are carried out at  tree level and the decays of the top quark and $W$ boson 
are treated in narrow-width approximation. 
The photon radiation can occur in both top quark
production and top quark decay. As a result, the following processes have to be considered:
\begin{eqnarray}
pp & \rightarrow & tj\gamma,~ t\rightarrow Wb \rightarrow l\nu b, \nonumber \\
pp & \rightarrow & tj, ~t\rightarrow Wb\gamma \rightarrow l\nu b \gamma, \nonumber \\
pp & \rightarrow& t j, t\rightarrow Wb,~W\rightarrow l\nu\gamma. 
\end{eqnarray}
The Feynman diagrams contributing to the single top plus photon production are depicted in
Fig.\ref{feynman}.

\begin{figure}[htb]
\begin{center}
\vspace{1cm}
\resizebox{0.45\textwidth}{!}{\includegraphics{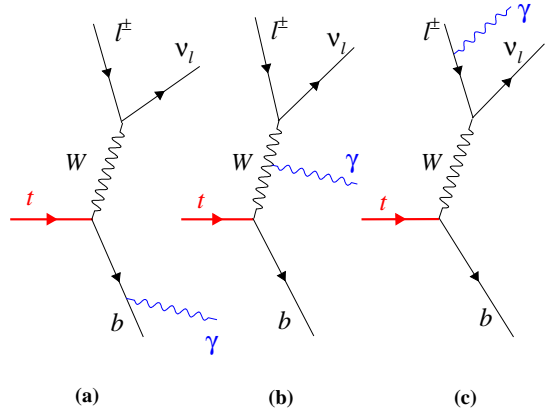}}  \caption{  Additional Feynman diagrams contributing to $l\nu b\gamma$ production in proton-proton collisions at the LHC.}\label{feynman1}
\end{center}
\end{figure}

The additional
Feynman diagrams corresponding to the cases that photon is emitted from the $W$ boson, b-quark and the charged lepton
are presented in Fig.\ref{feynman1}. 
In general, we cannot distinguish between the photon emission from top quark production and decay. 
As a result, the non-negligible interference effects between  these two types need to be  considered.

In order to perform  numerical calculations and simulations we have chosen the SM input parameters to
be: $m_{t}$ = 173.2 GeV, $m_{W} = 80.39$ GeV and  $G_{F} = 1.16639\times10^{-5}$ GeV$^{-2}$.
The event generation and cross section calculations are performed at leading-order with MadGraph 5 \cite{mg5,mg51}
including the spin correlations for the subsequent decays of the top quark.
We employ NNPDF3.0  \cite{nnpdf} parton distribution functions  and choose the value of the
factorization and renormalization scales event-by-event to be $\mu_{R} = \mu_{F} = Q_{0} = \sqrt{m_{t}^{2}+\sum_{i} p_{T}^{2}(i)}$,
where the sum is over the visible final state particles. All calculations are performed for proton-proton collisions
 at the center-of-mass energy of 13 TeV.

The cross section of process  $pp\rightarrow t+j+\gamma$  becomes divergent 
when the emitted photon is collinear to the initial particle. In order to avoid
such divergencies, we impose a minimum cut on the transverse momentum of the photon. 
To quantify the importance of contributions from the additional diagrams appearing in the 
top quark decays (Fig.\ref{feynman1}), we compare the cross sections of $pp \rightarrow tj\gamma\times Br(t\rightarrow \mu\nu b)$
and $pp\rightarrow tj\rightarrow \mu\nu b\gamma$.

\begin{figure}[htb]
\begin{center}
\vspace{1cm}
\resizebox{0.45\textwidth}{!}{\includegraphics{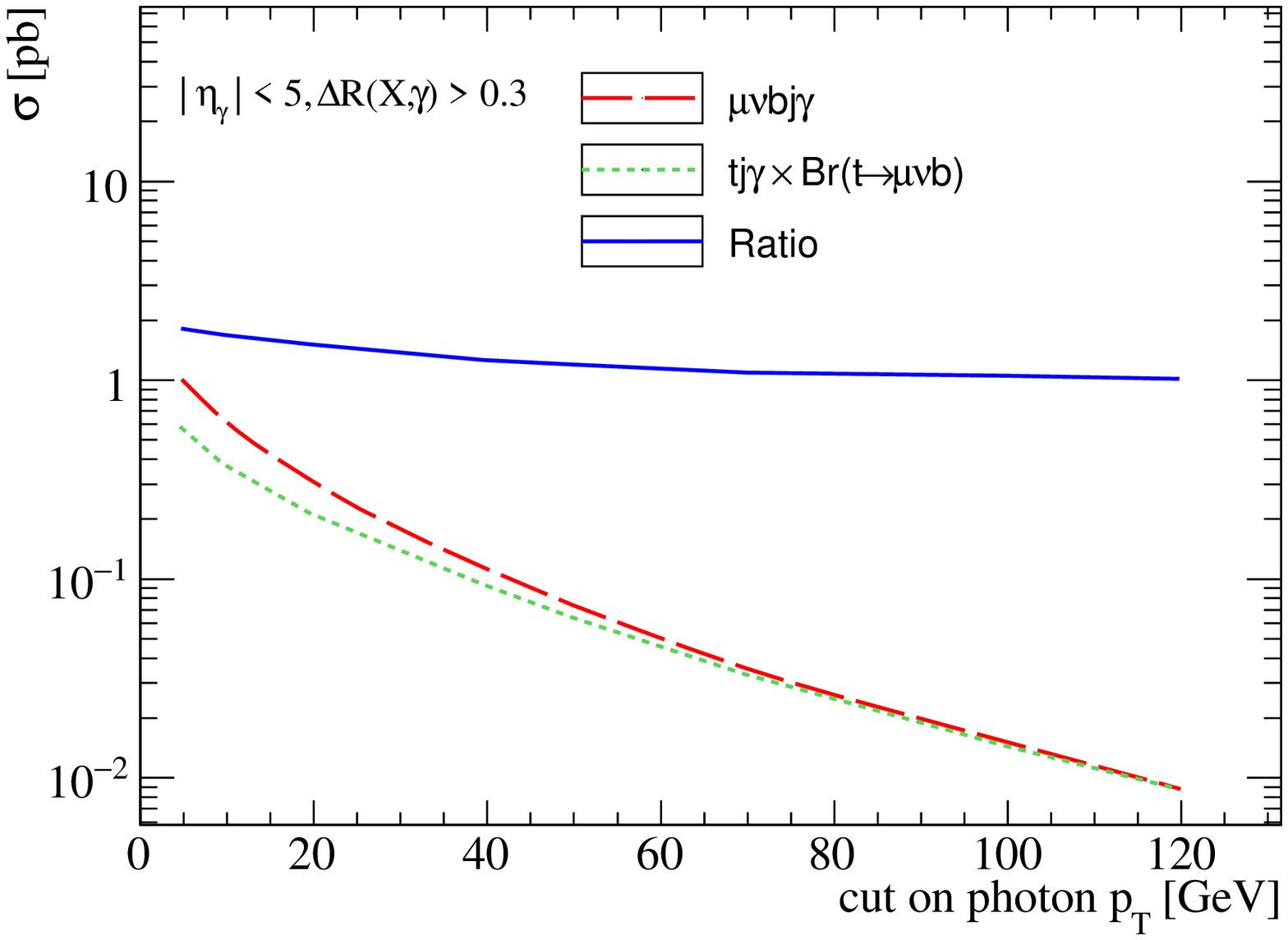}}  
\resizebox{0.45\textwidth}{!}{\includegraphics{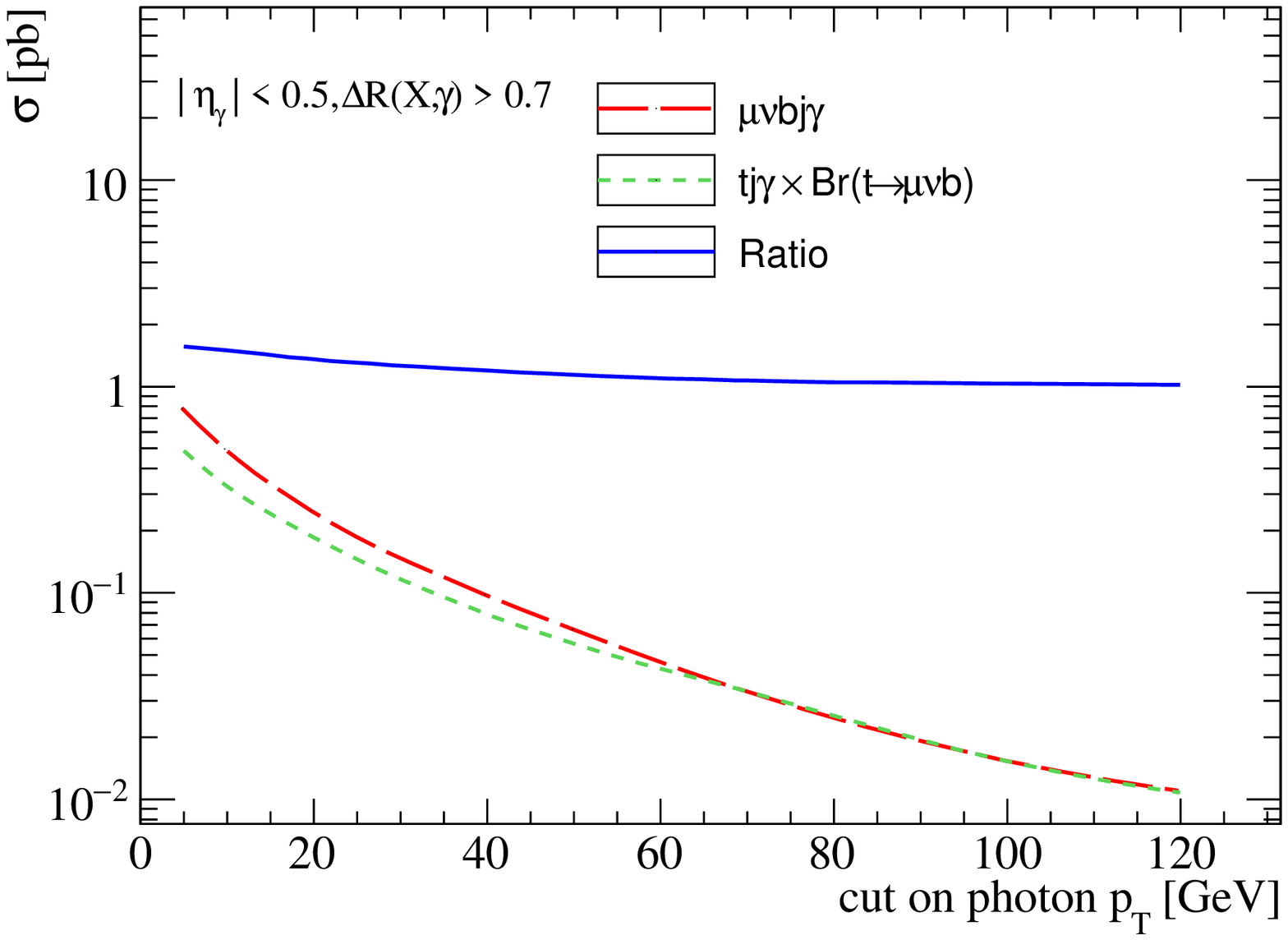}}  
\caption{  The cross section of $pp \rightarrow tj\gamma\times Br(t\rightarrow \mu\nu b)$
and $pp\rightarrow tj\rightarrow \mu\nu b\gamma$ as a function of cut on the photon transverse momentum and the ratio of 
the rates. The cross sections are presented with $\Delta R$ cuts of photon and other objects to be greater than $0.3$ (left) and $0.7$ (right).}\label{xsection37}
\end{center}
\end{figure}

In Fig.\ref{xsection37}, the cross sections of $pp \rightarrow tj\gamma\times Br(t\rightarrow \mu\nu b)$
and $pp\rightarrow tj\rightarrow \mu\nu b\gamma j$ are shown as a function of cut on the photon transverse momentum.
The ratio  $\sigma(\mu\nu b\gamma j)/\big(\sigma(tj\gamma) \times Br(t\rightarrow \mu\nu b)\big)$ is also calculated to  
show the importance of the new diagrams depicted in Fig.\ref{feynman1}.

The cross sections and ratio are shown for two cases of angular separation between the 
photon and all other final state objects $\Delta R(X,\gamma) > 0.3$ and $\Delta R(X,\gamma) > 0.7$, where 
$\Delta R(X,\gamma) = \sqrt{(\eta_{\gamma}-\eta_{X})^{2}+(\phi_{\gamma}-\phi_{X})^{2}}$.
As it can be seen,  including the contributions where the photon is emitted from
the decay products of the top quark leads to enhance the cross section by a factor
of 1.5 when a minimum cut of 20 GeV is applied on the photon transverse momentum. 
The magnitudes of the cross sections and ratio decrease with increasing the minimum cut on the
photon transverse momentum.  The amplitudes of the Feynman diagrams presented in 
Fig.\ref{feynman1} are suppressed when we increase the cut on photon transverse momentum
so that at a cut around 80 GeV, the cross sections are equal and the ratio tends to unity for $ \Delta R(X,\gamma) > 0.3$. 
By comparing the left and right plots in Fig.\ref{xsection37} we observe that the cut at which the ratio is equal to one
 depends on $\Delta R(X,\gamma)$ cut and it decreases with increasing the cut on $ \Delta R(X,\gamma)$.
 Applying a  cut of 0.7 on the angular separation between photon and other final state particles $ \Delta R(X,\gamma)$ leads to decrease 
 the value of minimum $p_{T}$ cut at which the contribution of the additional Feynman diagrams, presented in Fig.\ref{feynman1},
 are quite suppressed.

\section{Normalized cross section}
\label{xr}

The anomalous triple gauge boson couplings WW$\gamma$ and the anomalous top quark dipole moments t$\bar{\text{t}}\gamma$
contribute to the single top quark production in association with a photon at the LHC. In particular,
 diagram (c) in Fig.\ref{feynman} and  diagram (b) in Fig.\ref{feynman1} are affected by the
anomalous couplings WW$\gamma$. While the anomalous t$\bar{\text{t}}\gamma$ couplings
only contribute to the single top plus photon production via diagram (d) in Fig.\ref{feynman}.

In this section, we study the ratio between the production cross sections of
$tj\gamma$ and $tj$, $R=\sigma_{tj\gamma}/\sigma_{tj}$, versus the anomalous couplings $\kappa,\bar{\kappa},\Delta \kappa_{\gamma}$,
and $\lambda$ arising from the effective Lagrangians in Eq.\ref{eff1} and Eq.\ref{eff2}.
The advantage of using the ratio $R$ is to relieve from many experimental and theoretical sources of uncertainties with respect to the 
$tj\gamma$ production cross section. Experimental uncertainties such as
jet energy scale, lepton identification, b-jet tagging, and luminosity are canceled out. 
While  systematic uncertainties such as photon identification
and acceptance uncertainties are not dropped out completely.
The amount of theoretical uncertainty from the limited knowledge on parton distribution functions, variation of 
renormalization and factorization scales are significantly reduced in the ratio with respect to the total production cross section. 
As a result, in \cite{cdf} and \cite{cmsg}, the CDF and CMS collaborations have 
measured the ratio between the top quark pair production in association with a photon and the top pair
production rate.
In \cite{mangano}, the authors have shown that the top quark Yukawa coupling can be measured with 
an uncertainty of $1\%$ using the measured ratio of $\sigma_{t\bar{t}H}/\sigma_{t\bar{t}Z}$ in proton-proton 
collisions at the future circular collider FCC-hh. This can be achieved due to the cancellation of several sources 
of the systematic uncertainties. Also, in \cite{soreq} the authors make use of the ratio  $\sigma_{tt\gamma}/\sigma_{t\bar{t}Z}$
to constrain the top quark electroweak dipole moments and show that there is a significant reduction of uncertainties 
in this ratio.

Now, we turn to study the effects of the anomalous couplings
$\kappa$, $\bar{\kappa}$, $\Delta\kappa_{\gamma}$, and $\lambda$
on the normalized cross section $R=\sigma_{tj\gamma}/\sigma_{tj}$.
In order to perform the calculations and simulation, the
effective Lagrangians, Eq.\ref{eff1} and Eq.\ref{eff2}, are implemented
into the FeynRules \cite{feynrules} package and after that the model is exported to a Universal Feynrules Output (UFO) \cite{ufo} module which
is linked to MadGraph 5. 
Jets are reconstructed using the anti-k$_{t}$ \cite{antikt} algorithm and b-tagging 
efficiency of $60\%$ is assumed for tagging the jets originating from the hadronization of b-quarks.
We impose the following detector acceptance cuts on the transverse momentum, pseudorapidity, and 
angular separation:
\begin{eqnarray}
p_{T,\gamma} > 50~\text{GeV},~|\eta_{\gamma}| < 2.5~,~E_{T,\text{miss}} > 20 ~\text{GeV},~p_{T,l} > 20 ~\text{GeV},~|\eta_{l}| < 2.5, \nonumber \\
p_{T,j,b} > 20 ~\text{GeV},~\Delta R(m,n) > 0.4~(m\neq n),~|\eta_{b}|<2.5,~|\eta_{j}| <5.0,
\end{eqnarray}
where $m,n=\gamma,l,b,j$, $E_{T,\text{miss}}$ is the missing transverse energy
 and $\Delta R(m,n)$ is the separation between two particles $m$ and $n$ in the 
plane of pseudorapidity-azimuthal angle. 
In this study, no potential background processes are considered. 
The background processes to t-channel single top plus photon can be categorized into two classes:
 the irreducible and reducible background processes. The irreducible background process
 comes from the SM production of W$\gamma$+jets which has a similar final state to the signal process.
 The  reducible background processes originate from various SM processes that 
 have different final state from the signal but show similar signature to single top quark in association with a photon
 because of  misidentication of the final state objects. The main reducible background processes
 are W+jets and top pair events, with a jet misidentified as a photon.
There are background processes with electrons from the decays of W and/or Z boson 
which are misidentified as photons in the detector. Z+jets process is an example of this type 
of backgrounds. Negligible background contributions can come from processes with 
di-lepton in the final state (such as Z$(\rightarrow l^{-}l^{+})\gamma$+jets) where 
one of the  leptons is outside of detector coverage.

Since the SM prediction for the cross section of signal is small and there are many sources of
background processes, it is necessary to increase as much as possible the separation between
 signal and background events. This would lead to achieve a good sensitivity to the 
 anomalous couplings. In order to obtain the best discriminating power, 
 a multivariate classification based on boosted decision trees (BDT) or neural network (NN) 
 could be used \cite{bdt}.
 Further improvement on this study would be to consider the detector 
effects as well as all background processes  to have a more realistic estimate.
 
The leading order SM prediction for the normalized cross section $R$ is found to be $0.27\%$.
We check the robustness of $R$ against variations of the renormalization and factorization scales 
and also the parton distribution functions. The ratio $R$ is calculated once with doubling and once with halving 
the scales, i.e. $\mu_{R} = \mu_{F} = Q_{0}/2$ and $\mu_{R} = \mu_{F} = 2\times Q_{0}$. The changes on $R$ due to 
the variation of scales are found to be $+1.1\%$ and $-0.7\%$ corresponding to lowering and increasing the scales, respectively.
To examine the stability of the ratio $R$ versus the variations of the parton distribution functions, three independent 
PDFs of NNPDF3.0 \cite{nnpdf},  CTEQ6L1 \cite{cteq} and MRST  \cite{mrst} PDF sets are used to calculate the normalized cross section $R$.
The change of the central value of $R$ due to using different PDFs is found to be less than $1\%$ while 
the corresponding uncertainty on the total cross section of single top plus photon is around $3\%$.

In order to obtain the sensitivity on the anomalous couplings using the normalized cross section $R$,
we choose larger values than the uncertainties from the variation of renormalization and factorization scales
and PDF. The results are presented with two assumed uncertainties of $5\%$ and $10\%$ on measuring
the normalized cross section $R$. Assuming such uncertainties are meaningful given that the LHC
is going to deliver an anticipated integrated luminosity of around 300 fb$^{-1}$ in its Run 3 in which the statistical
uncertainties in single top quark and single top quark plus photon processes are subdominant.

\begin{figure}[htb]
\begin{center}
\vspace{1cm}
\resizebox{0.4\textwidth}{!}{\includegraphics{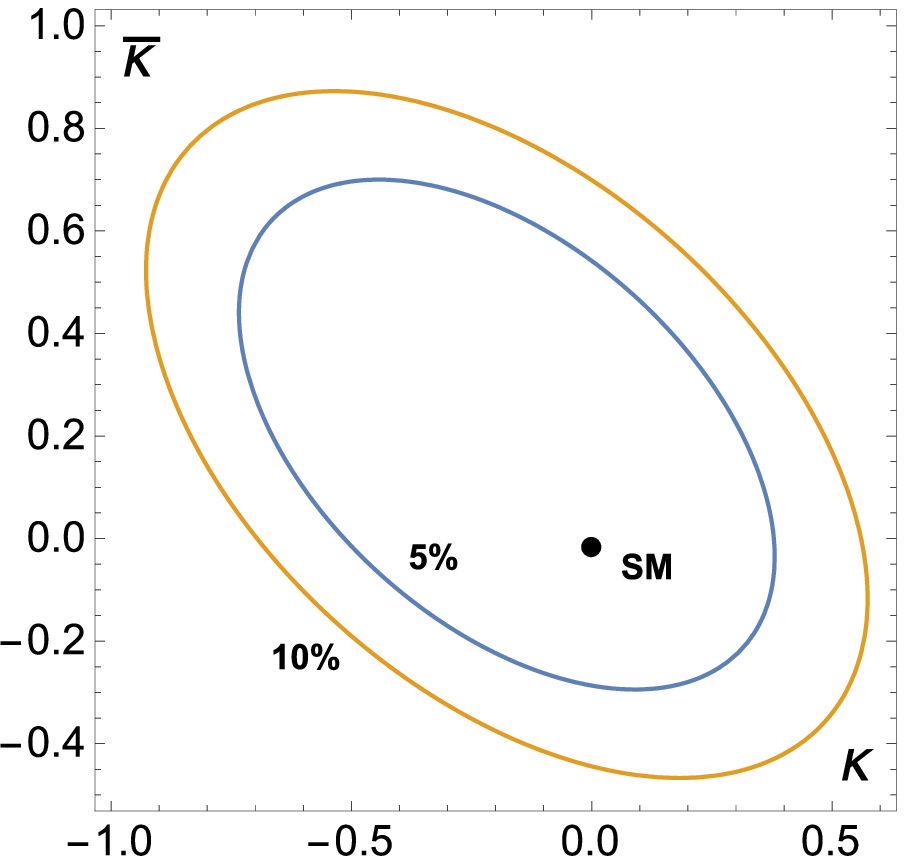}}  
\resizebox{0.4\textwidth}{!}{\includegraphics{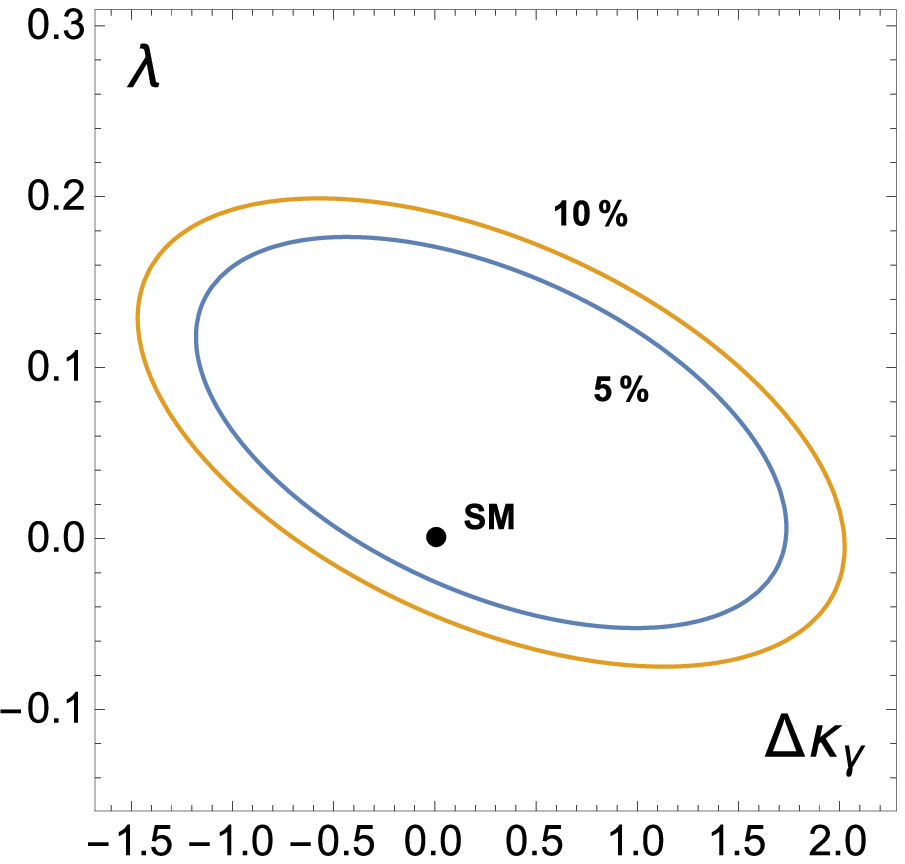}}  
\caption{  The $95\%$ CL contours in the plane of anomalous couplings ($\kappa,\bar{\kappa}$) (left panel) and ($\lambda,\Delta \kappa_{\gamma}$) (right panel) corresponding to measurement of cross section ratio $R=\sigma_{tj+\gamma}/\sigma_{tj}$ are presented with assumed uncertainties of $5\%$ and $10\%$ . }\label{ratio1}
\end{center}
\end{figure}

Fig.\ref{ratio1} shows the $95\%$ CL contours for the anomalous
top quark dipole couplings $\kappa$ and $\bar{\kappa}$ (left panel) and for the anomalous triple gauge 
couplings $\Delta\kappa_{\gamma}$ and $\lambda$ (right panel) with the assumed uncertainties of $5\%$ and $10\%$
on $R$ measurement.  With the uncertainty of $5\%$, the $95\%$ CL bounds on the couplings are found to be
 $\kappa\in[-0.72,0.38]$, $\bar{\kappa}\in[-0.27,0.67]$, $\lambda\in[-0.05,0.19]$
and $\Delta\kappa_{\gamma}\in[-1.1,1.8]$. The limits on the anomalous dipole moments
of the top quark $\kappa$ and  $\bar{\kappa}$ are corresponding to the following limits on the
electric and magnetic dipole moments of the top quark:
\begin{eqnarray}
a_{t} \in [-1.08,0.57] ~\text{and}~ d_{t} \in [-1.54,3.82]\times 10^{-17}e.\text{cm}.
\end{eqnarray}

For the electric and magnetic dipole moments, an improvement of around an order of magnitude is reachable in comparison with the
the constraints obtained from the combination of direct ($pp\rightarrow t\bar{t}\gamma$) and indirect ($b\rightarrow s\gamma$) 
searches mentioned previously. No considerable sensitivity is observed on the anomalous triple gauge 
boson coupling $\Delta\kappa_{\gamma}$ while the lower bound on $\lambda$ is comparable with the one obtained 
from the W$\gamma$ process.
In the next section, we suggest a particular angular asymmetry in single top quark production in association with 
a photon and examine its sensitivity to the anomalous couplings t$\bar{\text{t}}\gamma$ and WW$\gamma$.

\section{Angular asymmetry }
\label{asymmetry}

In this section, we construct an asymmetry from the 
kinematic observables of the final state particles of single top plus photon process
 to probe the anomalous t$\bar{\text{t}}\gamma$ and  WW$\gamma$ couplings.
The ability of this asymmetry to distinguish the contributions from the different
Lorentz structures in the vertices of t$\bar{\text{t}}\gamma$ and WW$\gamma$ due to their 
particular characteristic momentum dependence is also investigated. 

\begin{figure}[htb]
\begin{center}
\vspace{1cm}
\resizebox{0.45\textwidth}{!}{\includegraphics{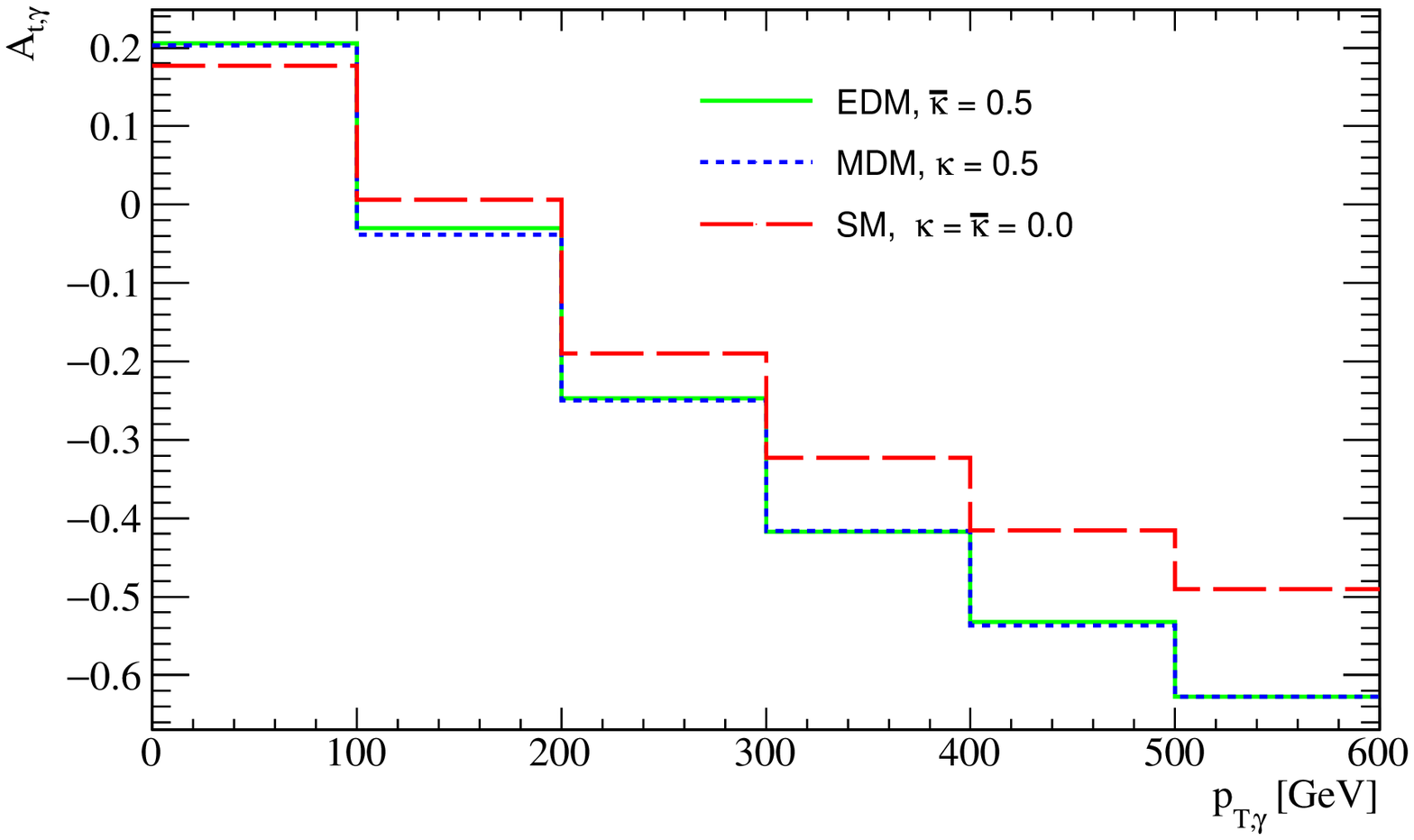}}  
\resizebox{0.45\textwidth}{!}{\includegraphics{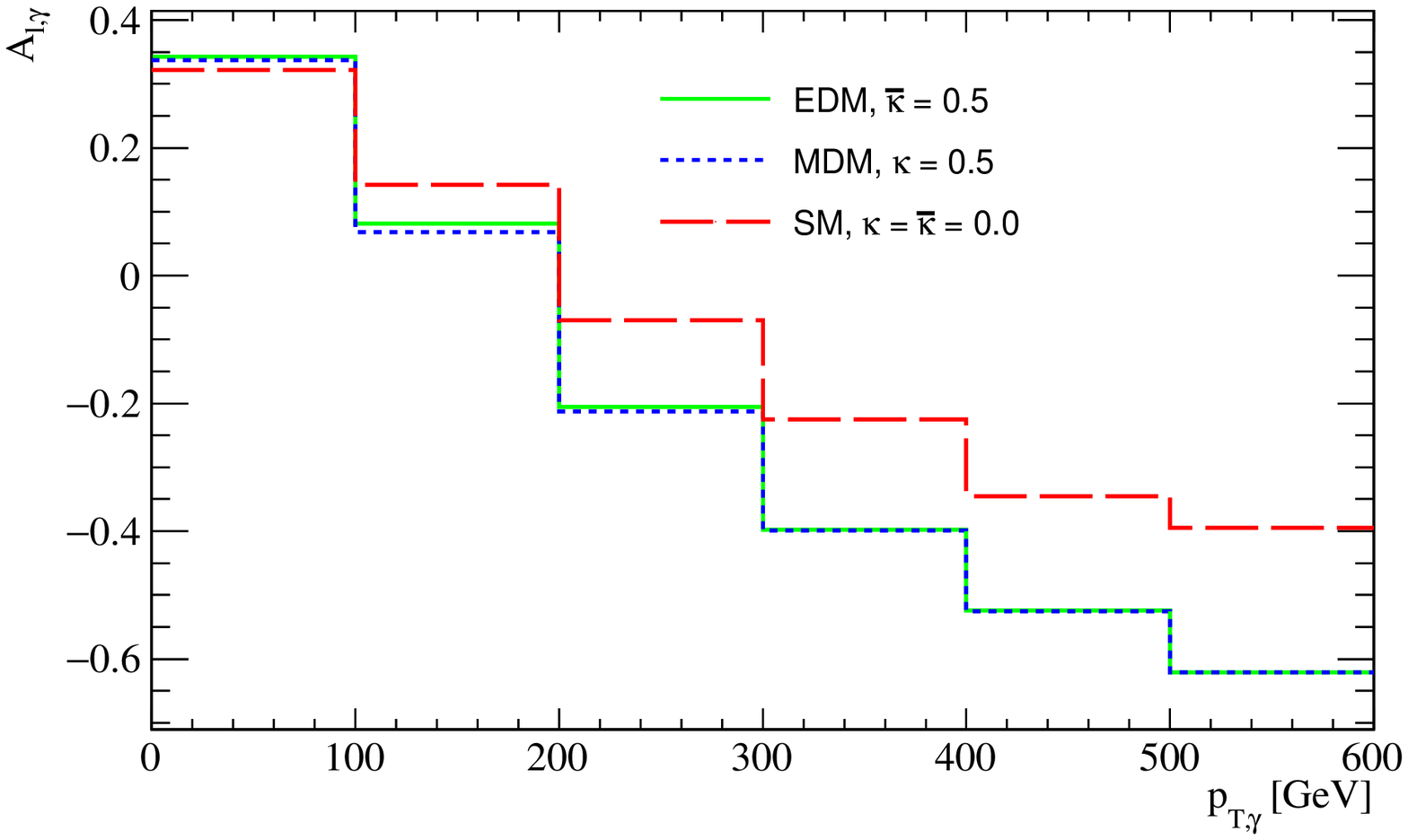}}  
\caption{  The dependence of the asymmetries  on the photon $p_{T}$. The left panel shows the
$A_{t,\gamma}(p_{T,\gamma})$  for the SM case and in the presence of top quark dipole moments while the $A_{l,\gamma}(p_{T,\gamma})$
is presented in the right panel. The dashed red curve depicts the SM case and the solid green and dashed blue curves show the asymmetries in the presence of $\kappa$ and $\bar{\kappa}$. }\label{asymmetry1}
\end{center}
\end{figure}

The presence of the anomalous t$\bar{\text{t}}\gamma$ is expected to affect the angular separation between the top quark and photon $\Delta R(t,\gamma)$ 
in single top quark production in association with a photon as well as other kinematic variables.
It is also expected that the anomalous couplings WW$\gamma$
modify the angular distribution of the emitted photon as there are contributions where the photon is radiated from 
the exchanged $W$ boson in both top quark production and decay. 
We consider the cosine of the angle between the top quark and photon,  $\cos\big(\vec{p}_{t},\vec{p}_{\gamma}\big)$,
to construct the following asymmetry observable:
\begin{eqnarray}
A_{t,\gamma} = \frac{N\big(\cos(\vec{p}_{t},\vec{p}_{\gamma}) > 0\big)- N\big(\cos(\vec{p}_{t},\vec{p}_{\gamma}) <0\big)}
{N\big(\cos(\vec{p}_{t},\vec{p}_{\gamma}) > 0\big)+N\big(\cos(\vec{p}_{t},\vec{p}_{\gamma}) <0\big)}, 
\end{eqnarray}
where $\vec{p}_{t}$ and $\vec{p}_{\gamma}$ are respectively, the momentum vector 
of the top quark and photon in the lab frame. 
In this work, we look at this asymmetry with respect to the photon $p_{T}$ and calculate 
it in different bins of the photon transverse momentum. We choose the photon transverse momentum 
because from the  experimental point of view it is a very clean object and easy to reconstruct.
The distribution $A_{t,\gamma}(p_{T,\gamma})$ is shown in the left panel of Fig.\ref{asymmetry1}.
The dashed red curve shows the behavior of $A_{t,\gamma}(p_{T,\gamma})$ 
in the SM while the solid green and dashed blue curves show the asymmetries in the presence of electric and magnetic 
dipole moments, respectively.

\begin{figure}[htb]
\begin{center}
\vspace{1cm}
\resizebox{0.55\textwidth}{!}{\includegraphics{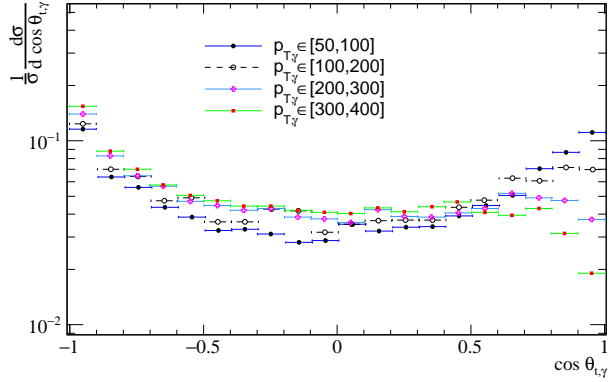}}  
\caption{ The normalized distribution of the cosine angle between the top quark and photon 
momenta in the lab frame in different bins of the photon transverse momentum predicted by the SM.  }\label{cosbin}
\end{center}
\end{figure}

The  qualitative  behavior  of   $A_{t,\gamma}(p_{T,\gamma})$ for the SM curve  can  be  understood  by looking at the
distribution of $\cos\big(\vec{p}_{t},\vec{p}_{\gamma}\big)$ in different bins of photon transverse momentum.
Fig.\ref{cosbin} shows the distributions of $\cos\big(\vec{p}_{t},\vec{p}_{\gamma}\big)$ for the cases that
$p_{T,\gamma}\in[50,100], ~[100,200],~[200,300],~[300,400]$. As it can be seen, 
photons with transverse momentum residing in the range of 50-100 GeV tend to be emitted 
mostly close to the top quark momentum direction. Going up to the higher momentum ranges 
leads increase the probability for the photons to be radiated at large angles with respect to the top quark.
This causes to have larger number of events with $\cos(\vec{p}_{t},\vec{p}_{\gamma}) < 0$ as
 large photon $p_{T}$ is  corresponding to emission with large angles with respect to 
the top quark. Higher photon transverse momentum is correlated with larger angles between the top and photon momenta.
Therefore, the events with very high $p_{T}$ photon mostly tend to have $\cos(\vec{p}_{t},\vec{p}_{\gamma}) < 0$.
This  causes $A_{t,\gamma}(p_{T,\gamma})$ to decrease with increasing the photon transverse momentum.

There are reasons which motivate to use the cosine of the angle between the 
charged lepton and the photon $\cos(\vec{p}_{l},\vec{p}_{\gamma})$ instead of
$\cos(\vec{p}_{t},\vec{p}_{\gamma})$ and consequently $A_{l,\gamma} $ as a reconstruction-independent asymmetry with the following 
definition instead of $A_{t,\gamma} $:
\begin{eqnarray}
A_{l,\gamma} = \frac{N\big(\cos(\vec{p}_{l},\vec{p}_{\gamma}) > 0\big)- N\big(\cos(\vec{p}_{l},\vec{p}_{\gamma}) <0\big)}
{N\big(\cos(\vec{p}_{l},\vec{p}_{\gamma}) > 0\big)+N\big(\cos(\vec{p}_{l},\vec{p}_{\gamma}) <0\big)},
\end{eqnarray}
where $\vec{p}_{l}$ is the momentum vector of the charged lepton.
The reasons that $A_{l,\gamma} $ is considered as an optimum observable with 
respect to $A_{t,\gamma} $ are as follows.
First, $A_{l,\gamma} $ as a reconstruction-independent quantity have no combinatorial issues, therefore
the sensitivity to the way of choosing the top decay products is significantly reduced. Second,  $A_{l,\gamma} $ 
is less sensitive to  modeling of the various distributions involved with respect to 
the $A_{t,\gamma} $  and consequently the related systematic uncertainties are under better control.  
The photon radiation coming from the top quark decay products changes the kinematics
and smears the relation between $A_{t,\gamma}(p_{T,\gamma})$ and $A_{l,\gamma}(p_{T,\gamma})$.
The behavior of $A_{l,\gamma}(p_{T,\gamma})$ is  depicted in the right panel of Fig.\ref{asymmetry1}.

\begin{figure}[htb]
\begin{center}
\vspace{1cm}
\resizebox{0.45\textwidth}{!}{\includegraphics{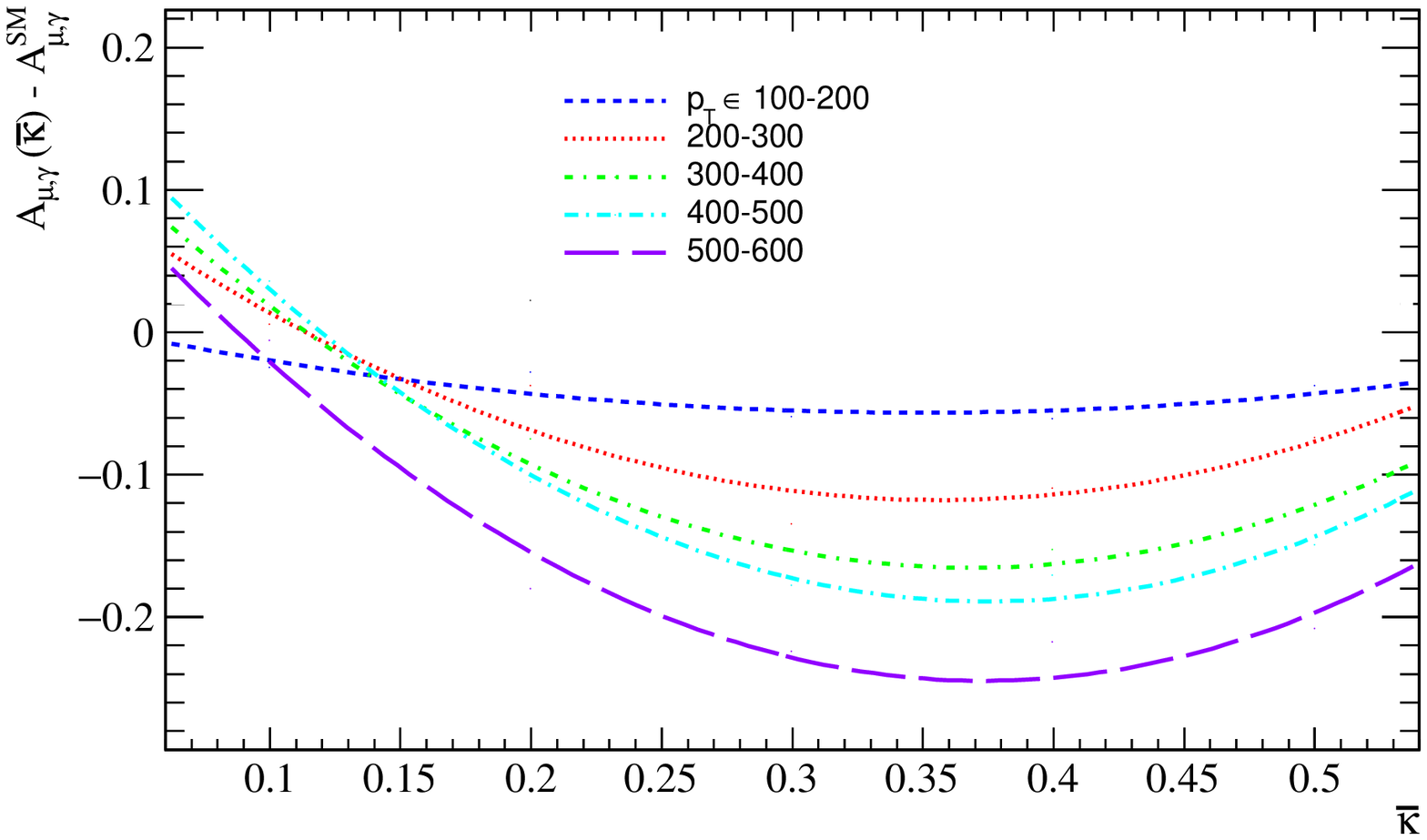}}  
\resizebox{0.45\textwidth}{!}{\includegraphics{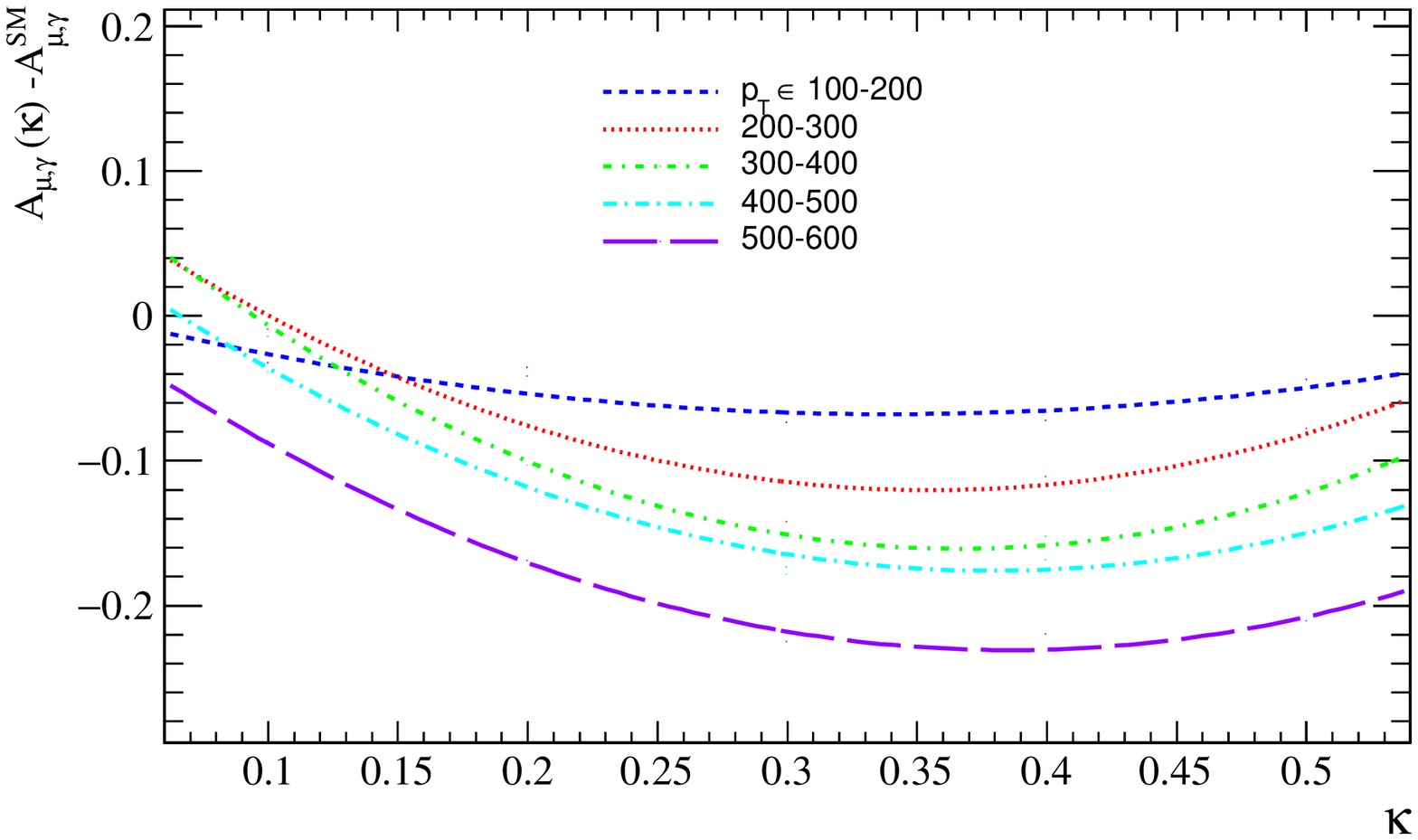}}  
\caption{  The dependence of difference of $A_{l,\gamma}(p_{T,\gamma})$ in the presence of the dipole moments from the SM value 
 in various photon $p_{T}$ bins in terms of $\bar{\kappa}$ (left panel) and $\kappa$ (right panel) is shown. }\label{bins}
\end{center}
\end{figure}

As it can be seen in Fig.\ref{asymmetry1}, in the events with large photon $p_{T}$ the presence of 
electric and magnetic dipole moments for the top quark reduces the asymmetries from the SM predictions. This allows to 
obtain the expected bounds on $\bar{\kappa}$ and $\kappa$.
To this purpose, first we find the dependence of $A_{l,\gamma}(p_{T,\gamma})$ in the $p_{T}$ bins
 of photon. In Fig.\ref{bins}, the dependence of the difference of $A_{l,\gamma}(p_{T,\gamma})$ in the presence 
 of top quark dipole moments from the SM value is presented in various photon $p_{T}$ bins in terms of 
 $\bar{\kappa}$ and $\kappa$.
As it is expected, when we go to larger photon transverse momentum, larger deviations from the SM prediction is 
observed.

In order to obtain the sensitivity of the anomalous couplings 
a $\chi^{2}$ analysis is performed, where the sum
of the variance of the asymmetry over all bins are computed.
In the presence of new couplings,  the $\chi^{2}$
is a function of  anomalous couplings $\kappa$ and $\bar{\kappa}$
and defined as:
\begin{eqnarray}
\chi^{2}\big(\kappa,\bar{\kappa}\big) = \sum_{i}^{n_{bin}}\Big(\frac{A_{l,\gamma}(\kappa,\bar{\kappa})[i]-A_{l,\gamma}^{\text{SM}}[i]}
{\Delta A_{l,\gamma}^{\text{SM}}[i]}\Big)^{2},
\end{eqnarray}
where $A_{l,\gamma}(\kappa,\bar{\kappa})[i]$ and $A_{l,\gamma}^{\text{SM}}[i]$ are the 
asymmetry predicted by the theory involving $\kappa$ and $\bar{\kappa}$ and the SM prediction for 
$i$th bin of photon transverse momentum. 
$\Delta A_{l,\gamma}^{\text{SM}}[i]$ represents all sources of 
the  uncertainties in $i$th bin of photon $p_{T}$.  
In this work, we only consider the statistical uncertainty which can be obtained using 
the following formula:
\begin{eqnarray}
\Delta A_{l,\gamma}^{\text{SM}}= \sqrt{\frac{1-(A_{l,\gamma}^{\text{SM}})^{2}}{\sigma_{\text{SM}}\times\mathcal{L}}},
\end{eqnarray}
where $\mathcal{L}$, $A_{\text{SM}}$, and $\sigma_{\text{SM}}$ 
are  the  integrated  luminosity,  the
value of the asymmetry and the cross section of the SM process, respectively.
We perform the $\chi^{2}$ analysis on $A_{l,\gamma}(\kappa)$ and  $A_{l,\gamma}(\bar{\kappa})$ distributions
shown in the right panel of Fig.\ref{asymmetry1} to extract the upper limits separately on the anomalous couplings
 $\kappa$ and $\bar{\kappa}$
at $95\%$ CL. The results are shown in Table \ref{res1} for two different integrated luminosities 30 fb$^{-1}$ and
300 fb$^{-1}$.

\begin{table}[htbp]
\centering
\caption{The  $95\%$ CL upper limits on the electric and magnetic dipole moments of the top quark obtained from  single top+$\gamma$
channel at the LHC with the center-of-mass energy of 13 TeV and for integrated luminosities of 30 and 300 fb$^{-1}$.}
\label{table1}
\begin{tabular}{|c|c|c|}
\hline 
Coupling & $\int\mathcal{L}dt = 30$ fb$^{-1}$ & $\int\mathcal{L}dt = 300$ fb$^{-1}$   \\  \hline 
$d_{t}$ [$10^{-17}e$.cm] (${\bar{\kappa}}$)    &    1.2 (0.21)  &    0.51 (0.09)       \\   
$a_{t}$     ($\kappa$)                                      &      0.43 (0.29)    &   0.16 (0.11)        \\   \hline  
\end{tabular}
\label{res1}
\end{table}

From Table \ref{res1}, we see that with 30 fb$^{-1}$ 
the top quark electric and magnetic dipole moments could be probed down to the order of $10^{-17}e.$cm 
and 0.43, respectively.
Using 300 fb$^{-1}$ integrated luminosity of data, the upper limit on the top quark magnetic dipole moment $a_{t}$ is found to be 0.16. 
This is still much larger than the SM prediction for $a_{t}$ which is 0.02.

\begin{figure}[htb]
\begin{center}
\vspace{1cm}
\resizebox{0.55\textwidth}{!}{\includegraphics{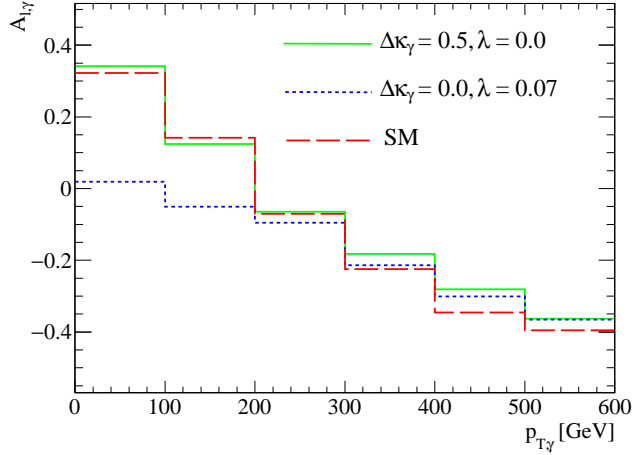}} 
\caption{  The dependence of the asymmetry $A_{l,\gamma}$ on the photon $p_{T}$. The plot shows 
$A_{l,\gamma}(p_{T,\gamma})$  for the SM case and in the presence of anomalous triple gauge 
boson coupling WW$\gamma$. The red dashed curve depicts the SM case and the solid green and dashed blue curves show the asymmetries in the presence of $\Delta \kappa_{\gamma}$ and $\lambda$. }\label{bins2}
\end{center}
\end{figure}

Now, we turn to study the sensitivity of the proposed asymmetry $A_{l,\gamma}$ to the anomalous triple gauge
boson coupling WW$\gamma$ as introduced by the Lagrangian in Eq.\ref{eff2}. The distribution of $A_{l,\gamma}$
as a function of photon $p_{T}$ is shown in Fig.\ref{bins2} for the SM, and for cases that anomalous WW$\gamma$
couplings are switched on. As it is expected the behavior of $A_{l,\gamma}(p_{T,\gamma})$ in the presence of 
$\kappa_{\gamma}$ is almost similar to the SM due to similarity in the couplings structure. However, the presence of 
anomalous coupling $\lambda$ distorts the shape of  $A_{l,\gamma}(p_{T,\gamma})$ in particular at photon transverse momentum 
smaller than 200 GeV. 
As no significant deviation from the SM in the presence of $\Delta \kappa_{\gamma}$ is observed, very loose sensitivity is expected to $\Delta \kappa_{\gamma}$. Following the same 
method as above leads to obtain upper limits on $\lambda$. 
The limits at $95\%$ CL on $\lambda$ are presented 
in Table \ref{res2}. Comparing to the current limits from the CMS and ATLAS experiments, the limits are loose however this could 
be a complementary study to W$\gamma$ channel for probing the anomalous triple gauge boson couplings.

\begin{table}[htbp]
\centering
\caption{The  $95\%$ CL upper limits on the
anomalous WW$\gamma$ couplings obtained from  single top plus $\gamma$
channel at the LHC with the center-of-mass energy of 13 TeV and for integrated luminosities of 30 and 300 fb$^{-1}$.}
\label{table1}
\begin{tabular}{|c|c|c|}
\hline 
Coupling & $\int\mathcal{L}dt = 30$ fb$^{-1}$ & $\int\mathcal{L}dt = 300$ fb$^{-1}$    \\  \hline 
$\lambda$                                           &    0.22      &     0.065              \\   \hline  
\end{tabular}
\label{res2}
\end{table}

It is notable that in addition to $\rm t\bar{t}\gamma$ and $\rm WW\gamma$ anomalous couplings,  single top plus photon production receives contributions from 
the anomalous Wtb vertex both in production and in decay. 
The most general effective Lagrangian describing the anomalous Wtb vertex has the following form \cite{eff2}:
\begin{eqnarray}
\mathcal{L}_{\rm Wtb} = -\frac{g}{\sqrt{2}}\bar{b}\gamma^{\mu}(V_{L}P_{L}+V_{R}P_{R})tW_{\mu}^{-}-
\frac{g}{\sqrt{2}}\bar{b}\frac{i\sigma_{\mu\nu}q^{\nu}}{m_{W}}(g_{L}P_{L}+g_{R}P_{R})tW_{\mu}^{-}+h.c.
\end{eqnarray}
where the coefficients $V_{L}$, $V_{R}$, $g_{L}$, and $g_{R}$ are dimensionless couplings. 
In the SM at tree level, $V_{L} = V_{tb}$ and other coefficients are equal to zero. 
 The existing bounds on these anomalous couplings from the weak radiative B-meson decay
 are \cite{bmeson}:  $-0.0007< V_{R} < 0.0025$, $-0.0013 < g_{L} < 0.0004$, and $  -0.15 <   g_{R}  < 0.57$. 
The constraints obtained at $95\%$ CL  on the anomalous
couplings  from W-boson helicities and t-channel cross section at the LHC
are \cite{wtblhc}: $-0.13 < V_{R}  < 0.18$, $-0.09 < g_{L}   < 0.06$, and $-0.15 < g_{R} < 0.01$. 
The normalized cross section $R = \sigma_{tj\gamma}/\sigma_{tj}$ introduced in Section \ref{xr}
is found to be almost insensitive to the anomalous Wtb couplings as the dependency is cancelled 
in the ratio. It is found that the variation of $g_{L,R}$ by an amount of 0.1 only leads to a change 
of around $0.1\%$ in $R$. The asymmetry observable $A_{l,\gamma}$ is found to be also 
insensitive to the anomalous Wtb couplings in  both shape and magnitude. It has a similar behavior to the SM
prediction in all bins of  
photon transverse momentum. As a result, $A_{l,\gamma}$ in single top production in association with a photon
is an angular observable which can distinguish only between possible new physics originating from
anomalous WW$\gamma$ interactions and top quark electric dipole moment.

%
%
\section{Summary and conclusions}\label{summary}

In this paper, we have investigated the possibility of measuring the non-standard couplings of 
t$\bar{\text{t}}\gamma$ and WW$\gamma$ through the process of single top quark production
in association with a photon at the LHC. 
Our analysis is based on the effective Lagrangian approach in which the modifications of t$\bar{\text{t}}\gamma$ and WW$\gamma$  interactions are coming from the dimension-six operators.
The analysis is carried out at leading order considering the 
processes in which the photon is either emitted in the production or from the  top quark decay products.

We examined the sensitivity of the ratio between the production rates of $tj\gamma$ and $tj$
to the anomalous t$\bar{\text{t}}\gamma$ and WW$\gamma$.
Many sources of the systematic uncertainties such as 
lepton and b-jet identification, jet energy scale,
and luminosity uncertainties almost cancel in the ratio. 
Experimental uncertainties like photon identification
and acceptance uncertainties are not cancelled completely in the ratio.
In particular, the leading order calculations
show that the systematic uncertainties originating from variations of scales and parton distribution
functions on ratio $R$ is are the level of around $1\%$.  
Based on assumed uncertainties of 5 and $10\%$ on measuring the normalized cross section $R$,
constraints on the anomalous triple gauge boson couplings WW$\gamma$ and t$\bar{\text{t}}\gamma$ are obtained.
The bounds on the anomalous top quark dipole moments with an assumed conservative 
uncertainty of $5\%$ are found to be $a_{t} \in [-1.08,0.57] ~\text{and}~ d_{t} \in [-1.54,3.82]\times 10^{-17}e.\text{cm}$.
We find that the cross section ratio $R$ has a weak dependence on the anomalous coupling $\Delta\kappa_{\gamma}$
and therefore loose bounds are obtained. However, strong lower bound  $-0.05$ on
another anomalous coupling $\lambda$  is reachable using the cross section ratio.

We also have defined a binned asymmetry observable using the distribution of 
the cosine angle between the charged lepton and photon as a tool 
to probe these new non-standard couplings. 
The asymmetry is calculated in the bins of the photon transverse momentum and has
a descending behavior with increasing the photon transverse momentum.
In our analysis, we have used a simple $\chi^{2}$ test in the absence of any systematic uncertainty
 to extract the sensitivity limits. Using the defined asymmetry, the sensitivity
bounds on the anomalous electric and magnetic dipole moments
can be significantly strengthened. With 300 fb$^{-1}$ of data, the limits are found to be
$|a_{t}| \leq 0.16 ~\text{and}~ |d_{t}| \leq 5.1 \times 10^{-18}e.\text{cm}$.
The proposed asymmetry is found to be  sensitive to only the anomalous 
gauge boson coupling $\lambda$ and no significant sensitivity to $\Delta\kappa_{\gamma}$ is seen.
An interesting observation is that the binned asymmetry has a discriminating capability 
between the SM and the anomalous couplings WW$\gamma$ and t$\bar{\text{t}}\gamma$ at 
the photon transverse momentum less than 200 GeV.
Further improvements could be achieved  including the higher order corrections to the
single top plus photon process in the presence of the anomalous couplings, considering the background processes and detector effects.
\\
%
{\bf Acknowledgments:}
Sara Khatibi is supported by Iranian National Science Foundation (INSF).
%

%
%

\end{document}